\documentclass{pasj00}

\begin{document}
\SetRunningHead{D. Nogami et al.}{The 2004 superoutburst of 2003aw}

\Received{2004/00/00}
\Accepted{2004/00/00}

\title{The Peculiar 2004 Superoutburst in the Helium Dwarf Nova, 2003aw}

\author{Daisaku \textsc{Nogami}$^1$, Berto \textsc{Monard}$^2$, Alon
\textsc{Retter}$^3$, Alex \textsc{Liu}$^4$, Makoto \textsc{Uemura}$^5$,
Ryoko \textsc{Ishioka}$^6$,}
\author{Akira \textsc{Imada}$^6$, and Taichi \textsc{Kato}$^6$}
\affil{$^1$Hida Observatory, Kyoto University,
       Kamitakara, Gifu 506-1314, Japan (nogami@kwasan.kyoto-u.ac.jp)}
\affil{$^2$Bronberg Observatory, PO Box 11426, Tiegerpoort 0056, South
Africa}
\affil{$^3$Department of Astronomy and Astrophysics, Pennsylvania State
University, 525 Davey Lab, University Part, \\
PA, 16802-6305, USA}
\affil{$^4$Norcape Observatory, PO Box 300, Exmouth 6707, Australia}
\affil{$^5$Departamento de F\'{\i}sica, Facultad de Ciencias
F\'{\i}sicas y Matem\'aticas, Universidad de Concepci\'on, Casilla
160-C,\\
Concepci\'on, Chile}
\affil{$^6$Department of Astronomy, Faculty of Science, Kyoto University,
Sakyo-ku, Kyoto 606-8502, Japan}


\KeyWords{accretion, accretion disks
          --- stars: dwarf novae
          --- stars: individual (SN 2003aw)
          --- stars: novae, cataclysmic variables
          --- stars: oscillations}

\maketitle

\begin{abstract}

 We conducted a time-resolved photometric campaign of the helium dwarf
 nova, 2003aw in 2004 May--June.  2003aw stayed at 14.7--15.7 mag for
 the first several days during this campaign, which is the plateau phase
 of this superoutburst.  This variable then rapidly decayed to
 $\sim$18.0 mag, still brighter by about 2 mag than its quiescence
 magnitude, and maintained this brightness for about 20 days, having
 short flares of $\sim$2 mag.  A long fading tail followed it.  We
 detected superhumps with a period of 0.02357(4) d [= 2036(3) s] during
 the plateau phase.  The whole light curve of the superoutburst in
 2003aw, taking into account the present data and those in the
 literature, perfectly resembles that of the 1996-1997 superoutburst of
 the peculiar WZ Sge-type hydrogen-rich dwarf nova, EG Cnc.

\end{abstract}

\section{Introduction}

Interacting binaries consisting of two white dwarfs (IBWDs) are called
AM CVn stars \citep{sol95amcvnreview,war95amcvn}.  Helium gas transfered
from the less massive white dwarf forms an accretion disk around the
primary.  There are 11 confirmed members including 2003aw in this class,
10 of which (and 2 candidates) are listed in table 1 in
\citet{wou03sn2003aw}, and SDSS J1240$-$01 is the most recently
discovered \citep{roe04sdssj1240}.  They have three activity groups,
mainly depending on the mass-transfer rate $\dot{M}$, like hydrogen-rich
cataclysmic variables (CVs; for a review, \cite{war95book}).  They are,
from the high-$\dot{M}$ systems, 1) stable systems with a hot accretion
disk, corresponding to nova-likes in CVs, 2) systems having outbursts,
corresponding to dwarf novae, 3) stable systems with a cold disk.
2003aw is the 6th member of the helium dwarf novae.

This star was discovered as a supernova (therefore named 2003aw) by
\citet{woo03sn2003avaw} from the unfiltered NEAT image taken with the
Palomar 1.2m Schmidt telescope.  The magnitude at the discovery on 2003
February 6 was 17.8, and this `supernova' was seen also in the images
taken on 2003 February 10 and 19.  \citet{cho03sn2003avaw} obtained an
optical spectrum, which had a quite blue continuum with weak He
\textsc{i} emission lines and Ca \textsc{ii} H and K absorption at
nearly zero redshift.  They commented that 2003aw is not a supernova,
but rather is a hydrogen-deficient dwarf nova similar to that proposed
for 1998di = KL Dra (see \cite{jha98kldraiauc6983}).

\citet{wou03sn2003aw} carried out intensive high-speed photometry of
2003aw in an active phase between 2003 February 28 and June 9.
They devided their observations into three states.  In the high state,
2003aw was in $V=17.6-19.0$, and gave rise to a flare reaching $V=16.5$.
In the intermediate state and quiescence, 2003aw was at about $V=19.6$
and at around $V=20.3$, respectively.  Several curious features were
revealed by them, such as superhumps with a period of 2041.5 ($\pm$0.3)
s, brightness cycles over 0.4 mag with a period of $\sim$16 h during the
high and intermediate state, and sidebands to the principal frequencies
having a constant frequency difference from the superhump harmonics.

Being informed by P. Woudt on 2004 May 17 that 2003aw was in outburst
at $V\sim15$ mag, which is at least 1.4 mag brighter than the maximum
during the 2003 February/March outburst [vsnet-alert 8131, see
\citet{VSNET} on VSNET], a photometric observation campaign was
immediately started by the VSNET collaboration team.  We report the
results in this paper.

\section{Observation}

\begin{table*}
\caption{Log of observations.}\label{tab:log}
\begin{center}
\begin{tabular}{clrccccccc}
\hline\hline
\multicolumn{3}{c}{Date} & HJD-2453000 & Exposure & Useful  & Mean
 & Comparison & Site$^{\dagger}$ & Sky  \\
         & &             & Start--End & Time (s) & Frames & Mag. &
 Star$^*$ & & Condition \\
\hline
2004 & May    & 18 & 143.960--144.058 & 60  &  86 & 15.7(0.1) & 1 & A \\
     &        &    & 144.005--144.019 & 30  &  35 & 15.2(0.8) & 1 & B & thin
 clouds \\
     &        &    & 144.173--144.329 & 28  & 440 & 15.2(0.1) & 1 & C \\
     &        & 19 & 145.179--145.304 & 28  & 353 & 15.3(0.1) & 1 & C \\
     &        & 20 & 145.953--146.060 & 60  &  93 & 14.7(0.1) & 1 & A \\
     &        &    & 146.182--146.287 & 28  & 202 & 15.7(0.1) & 1 & C \\
     &        & 22 & 147.956--148.057 & 60  &  90 & 16.6(0.2) & 1 & A \\
     &        & 24 & 149.974--149.979 & 30  &   4 & 16.1(0.7) & 1 & B & thin
 clouds \\
     &        & 25 & 150.960--151.057 & 60  &  61 & 17.6(0.3) & 1 & A \\
     &        & 26 & 152.169--152.260 & 28  & 254 & 16.4(0.2) & 1 & C \\
     &        & 27 & 152.975--153.051 & 90  &  50 & 18.3(0.3) & 2 & A \\
     &        & 28 & 153.959--154.068 & 90  &  72 & 18.1(0.3) & 2 & A \\
     &        & 29 & 154.953--155.057 & 90  &  66 & 18.1(0.3) & 2 & A \\
     & June   &  5 & 162.447--162.454 & 30  &   8 & 18.6(0.9) & 1 & D \\
     &        &  6 & 163.457--163.463 & 30  &   9 & 18.1(0.6) & 1 & D \\
     &        &  9 & 166.181--166.203 & 28  &  59 & 16.1(0.1) & 1 & C \\
     &        &    & 166.432--166.536 & 30  &  81 & 17.0(0.6) & 1 & D & thin
 clouds \\
     &        & 12 & 169.445--169.506 & 30  &  24 & 17.5(0.7) & 1 & D & thin
 clouds \\
     &        & 29 & 186.442--186.447 & 30  &   3 & 18.0(0.9) & 1 & D & thin
 clouds \\
\hline
\multicolumn{10}{l}{$^*$1: 2UCAC 29829807, 11.3 mag, 2: 2UCAC 29829819,
 16.1 mag}\\
\multicolumn{10}{l}{$^{\dagger}$A: 30cm Tel. + SBIG ST-7E (Exmouth,
 Australia), B: 30cm Tel. + SBIG ST-7E (Kyoto, Japan), C: 32cm} \\
\multicolumn{10}{l}{Tel. + SBIG ST-7E (Tiegerpoort, South
 Africa), D: 30cm Tel. + Pictor 416XT (Concepcion, Chile)} \\
 \end{tabular}
\end{center}
\end{table*}

The observations were carried out at four sites, Tiegerpoort in South
Africa (BM), Exmouth in Australia (AL), Concepcion in Chile (MU), and
Kyoto in Japan (RI, AM, and TK).  Table \ref{tab:log} summarizes the log
of the observations and the instruments.

The differential magnitudes of 2003aw were measured, assuming the
unfiltered CCD magnitudes of the nearby comparison stars, 2UCAC 29829807
and 2UCAC 29829819, to be 11.3 and 16.1, respectively, based on
the UCAC2 magnitude (see \cite{zac04UCAC2} regarding UCAC2).  The UCAC
bandpass is 579--642 nm, between $V$ and $R_{\rm c}$. The error in the
comparison magnitude is expected to be within 0.5 mag.

Heliocentric corrections to the observation times were applied before
the following analysis.

\begin{figure}
  \begin{center}
    \FigureFile(84mm,84mm){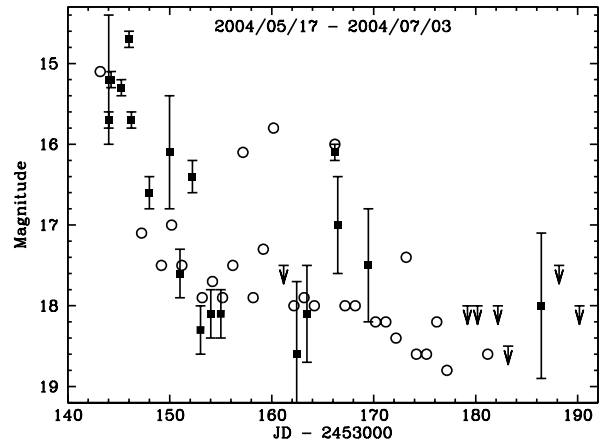}
  \end{center}
 \caption{Long-term light curve of the 2004 May outburst of 2003aw.
 The abscissa is HJD $-$ 2453000, and the ordinate is CCD magnitude.
 The open circles are the CCD observations reported to VSNET.  The error
 of these observations depends on the target brightness and the sky
 condition, but is smaller than 0.4 mag.  The filled squares are average
 magnitudes of each run of our data.  A quite bright state was recorded
 for several days at the start of our campaign.  Then 2003aw stayed
 around 18 mag for about 20 days, occasionally having sudden, short
 flares.  After this state, 2003aw gradually went down to quiescence.
 } \label{fig:lc}
\end{figure}

\section{Results}

The long-term light curve is drawn in figure \ref{fig:lc}.  At the first
stage of our campaign, 2003aw was recorded to be at 15.1--15.7 mag with
a brightening to 14.7 mag on May 20.  Even considering the zero-point
uncertainty of the magnitude and the color, this maximum magnitude far
exceeds the brightest magnitude at a brightening ($V=16.5$) in the
`high' state during the 2003 February/March outburst
\citep{wou03sn2003aw}.  The detection of the superhumps (described
later) indicates that this phase corresponds to the plateau phase of the
superoutburst in AM CVn stars and SU UMa-type dwarf novae.  It is
unclear how long the plateau phase lasted before our observations
started.

\begin{figure*}
  \begin{center}
    \FigureFile(168mm,168mm){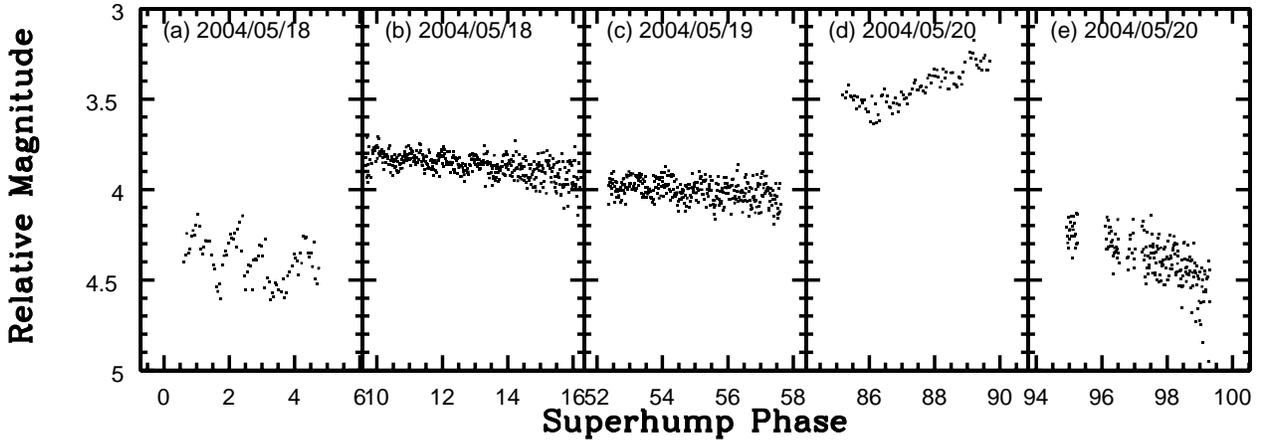}
  \end{center}
 \caption{Short-term light curves of the long runs during the plateau
 phase.  Stable superhumps with a small amplitude are seen in panels (b)
 and (c).  The superhump phase of the abscissa is adjusted using the
 data shown in these panels.  There are large amplitude ($\sim$0.5 mag)
 modulations in panel (a), which are possibly related to the
 superhumps.  There is no clear evidence of the superhump in panels (d)
 and (e).
 } \label{fig:sh}
\end{figure*}

After the decline from the plateau phase, 2003aw kept its brightness
around 18.0 mag, besides occasional flares reaching $\sim$ 15.8 mag, for
$\sim20$ days.  During this phase, 2003aw was still brighter by about
2.0 mag than in quiescence, which corresponds to the high state in
\citet{wou03sn2003aw}.  Judging from our observations and those reported
to VSNET (see figure \ref{fig:lc}), the duration of the flares is 1 day,
or so, but a periodicity of the flare can not be supported nor rejected
from the available sparse data.  This behavior is completely consistent
with that during the high state in the 2003 February/March active phase
\citep{wou03sn2003aw}.

A gradual fading to quiescence [the intermediate state defined by
\citet{wou03sn2003aw}] followed the high state.  We could not, however,
continue the observations well during and after this state with our 30cm
telescopes.

Figure \ref{fig:sh} shows the enlarged light curves of each long run
during the plateau phase.  There exist modulations with variable shapes,
amplitudes, and timescales.  Those shown in figures \ref{fig:sh}a have
a large amplitude up to 0.5 mag and an unstable timescale.  We see
small, stable modulations in figures \ref{fig:sh}b and \ref{fig:sh}c,
which are interpreted as superhumps.  A rapid brightening with a rate of
4.4 mag d$^{-1}$ is displayed in figure \ref{fig:sh}d, but 2003aw
decayed by 0.9 mag during the following 3 hours by the start of the run
of figure \ref{fig:sh}e.

\begin{figure}
  \begin{center}
    \FigureFile(84mm,84mm){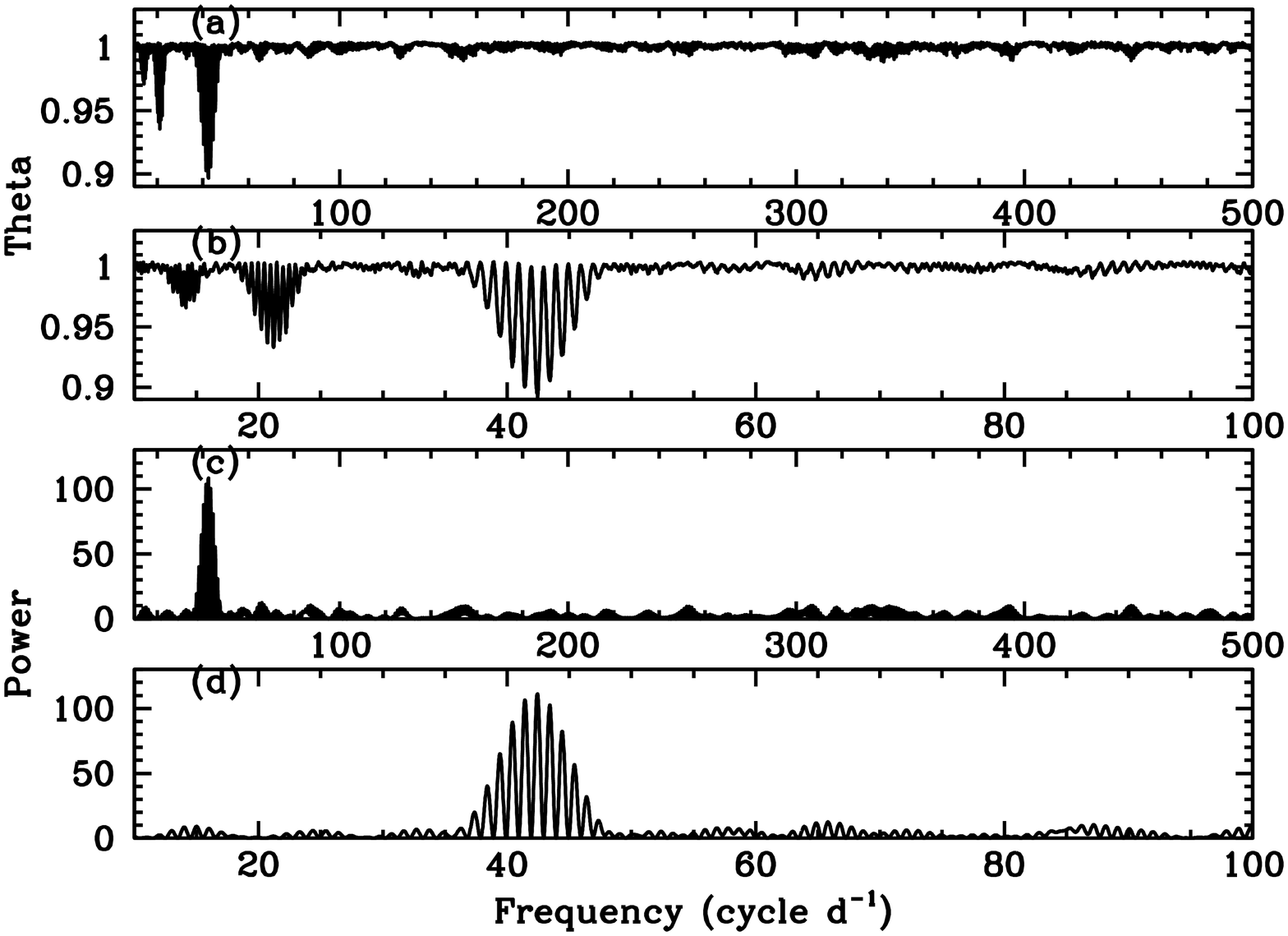}
  \end{center}
 \caption{Theta diagram of the PDM analysis and power spectrum,
 using the data shown in figure \ref{fig:sh}b and \ref{fig:sh}c, after
 removing variations with timescales longer than hours.  Panels (b)
 and (d) show the results in panels (a) and (c), respectively, zoomed on
 the frequency range around the superhump frequency.  Both analyses
 indicate the superhump period of 0.02357(4) d [= 42.44(7) cycle
 d$^{-1}$ = 2036(3) s].
 } \label{fig:pdm}
\end{figure}

We performed period analyses by the Phase Dispersion Minimization
(PDM) method \citep{PDM} and the Fourier transformation for all the data
in figures \ref{fig:sh}b and \ref{fig:sh}c containing the stable
superhumps.  Before these analyses, a pre-whitening removing variations
with timescales longer than hours was processed.  The resultant
$\Theta$ diagram and power spectrum are exhibited in figure
\ref{fig:pdm}.  The best estimated superhump period is 0.02357
($\pm$0.00004) d [= 2036 ($\pm$3) s].  The error of the period was
estimated using the Lafler-Kinman class of methods, as applied by
\citet{fer89error}.  This period is different by $\sim$2 $\sigma$ from
the value [2041.5 ($\pm$0.3) s] derived by \citet{wou03sn2003aw}.  This
difference can not be considered statistically significant, though a
similar difference in the superhump period in the different outburst
states were observed in another AM CVn star, V803 Cen
\citep{kat04v803cen}.  Including the data shown in figures \ref{fig:sh}a
and \ref{fig:sh}d did not improve the period determination.

We could not detect firm evidence of any sidebands, which would be (at
least partly) due to short coverages and low signal-to-noise ratios of
our data.

\begin{figure}
  \begin{center}
    \FigureFile(84mm,84mm){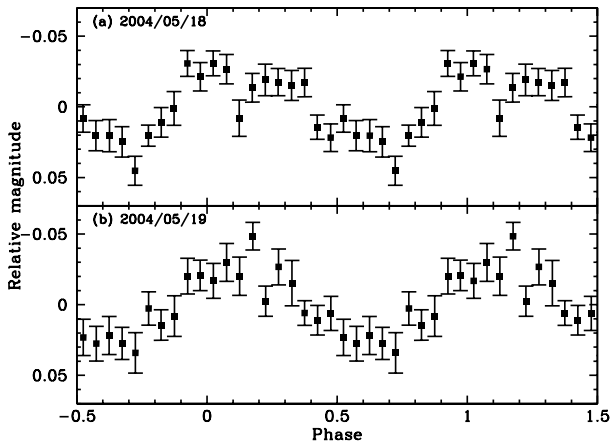}
  \end{center}
 \caption{The superhump profile folded with the superhump
 period derived in this work and averaged.  Panels (a) and (b) use the
 data shown in figures \ref{fig:sh}(b) and \ref{fig:sh}(c),
 respectively.  There is a hint of a dip around phase 0.1 in (a)
 superposed on a typical superhump shape with a rapid rise and a slow
 decay.  During the following $\sim$40 superhump cycles, the superhump
 shape became rather symmetric.
 } \label{fig:shshape}
\end{figure}

The averaged superhump profiles are exhibited in figure
\ref{fig:shshape}.  There is a hint of a dip after the superhump maximum
on 2004 May 18 (figure \ref{fig:shshape}a).  The same phenomenon was
observed by \citet{wou03sn2003aw} during the high state.  On this day,
the superhumps had a shape consisting of a rapid rise and a slow
decline, which is typically seen in hydrogen-rich SU UMa-type dwarf
novae.  On the next day, 2004 May 19, the superhump had evolved into
a symmetric shape (figure \ref{fig:shshape}b).

The timings of the superhump maxima were extracted by fitting the
average superhump light curve in figure \ref{fig:shshape}, and we tried
to check the change of the superhump period between HJD 2454144.173 and
2453145.304.  The superhump period, however, stayed constant during this
period.

A periodicity attributable to the orbital modulation could not detected
from any of our runs.

\section{Discussion}

Our observations of 2003aw revealed the existence of the plateau phase
of the superoutburst, which is brighter by $\sim$5.0 mag than its
quiescent magnitude, and the presence of superhumps with a period of
2036(3) s during this phase.  Following this phase, the star sharply
decayed with a rate over 1 mag d$^{-1}$ and stayed around 18.0 mag in
the high state for $\sim$20 days, having short flares with amplitudes
of $\sim$2.0 mag.  This variable gradually faded after that.
\citet{wou03sn2003aw} are considered to have missed the plateau phase,
and have steadily observed the high state and the fading tail.

The whole outburst light curve of 2003aw, taking into account the
observations by us and \citet{wou03sn2003aw}, has strong resemblance
to the superoutburst of the peculiar WZ Sge-type dwarf nova, EG Cnc, in
1996--1997 \citep{pat98egcnc,kat04egcnc}, in terms of the first quite
bright state (the plateau phase in 2003aw, and the main superoutburst
in EG Cnc), the following repetitive-outburst state still brighter than
the quiescence (the high state in 2003aw, and the rebrightening phase in
EG Cnc), and the long fading tail to the quiescence (the intermediate
state in 2003aw, and the long fading tail in EG Cnc). Superhumps were
observed throughout these states in both stars.

The thermal-tidal disk-instability models for the AM CVn stars have been
developed by \citet{sma83HeCV}, \citet{can84DIdegeneratebinary}, and
\citet{tsu97amcvn}.  To explain the peculiar outburst behavior in EG
Cnc, a modification of the disk-instability model for hydrogen-rich
dwarf novae was introduced by \citet{osa01egcnc}, adding an idea of the
viscosity decay in the cold disk.  A similar idea may be necessary in
reconstructing the full outburst property of 2003aw, although we need
further observations to clarify the total outburst pattern including
that outside the superoutbursts\footnote{Note that the outburst pattern
in AM CVn stars can vary in a timescale of years
\citep{pat00v803cen,kat01crboo,kat04v803cen}.}.  The repetition of
the large flares after the plateau phase will be a key also in trying
to consider the behavior of 2003aw with the models making use of the
variation in the mass-transfer rate \citep{war95amcvnproc}, or the
effect of the magnetic fields on the primary white dwarf
\citep{wou03sn2003aw}.

The repetitive-outburst (high) state casts another problem concerning a
notable difference in the superoutburst between 2003aw and EG Cnc.  In
2003aw, small brightness variations with an amplitude of $\ge$0.4 mag
and a period of $\sim$16 h were observed, other than a large flare of
1.9 mag by \citet{wou03sn2003aw}, while corresponding modulations were
not reported in EG Cnc \citep{pat98egcnc,kat04egcnc}.  The resemblance
of the outburst-light curves between an AM CVn star and a WZ Sge star
was first pointed out by \citet{kat04v803cen} concerning V803 Cen and WZ
Sge.  After the initial, long outburst stage, V803 Cen showed $\sim$1
mag oscillations with periods of 0.8--1.0 d, which resembles the
rebrightening stage of the 2001 superoutburst in WZ Sge.
\citet{kat04v803cen} attributed those oscillations to some sort of
thermal disk-instability, instead of the full-disk outbursts.  To fully
understand the outburst mechanism in 2003aw, we must explain the
0.4-mag oscillations with a timescale of 16 h and the occasional flares
up to 2 mag.

\vskip 3mm

The authors are thankful to amateur observers for continuously
reporting their valuable observations to the VSNET.  We are indebted to
P. A.  Woudt for his notification of the outburst and useful comments on
the draft as the referee.  This work is supported by Research
Fellowships of Japan Society for the Promotion of Science for Young
Scientists (MU and RI), and Grants-in-Aid for the 21st Century COE
``Center for Diversity and Universality in Physics'' from the Ministry
of Education, Culture, Sports, Science and Technology (MEXT) of Japan,
and also by Grants-in-Aid from MEXT (No. 13640239, 15037205).


\begin{thebibliography}{}

\bibitem[Cannizzo(1984)]{can84DIdegeneratebinary}
  Cannizzo, J.~K.\ 1984, \nat, 311, 443

\bibitem[Chornock, Fillipenko(2003)]{cho03sn2003avaw}
  Chornock, R., \& Fillipenko, A.~V.\ 2003, \iaucirc, 8084

\bibitem[Fernie(1989)]{fer89error}
  Fernie, J.~D.\ 1989, \pasp, 101, 225

\bibitem[Jha et~al.(1998)]{jha98kldraiauc6983}
  Jha, S., Garnavich, P., Challis, P., Kirshner, R., \& Berlind, P.\ 1998,
  \iaucirc, 6983

\bibitem[Kato et~al.(2004a)]{kat04egcnc}
  Kato, T., Nogami, D., Matsumoto, K., \& Baba, H.\ 2004a, \pasj, 56, S109

\bibitem[Kato et~al.(2001)]{kat01crboo}
  Kato, T. {et~al.}\ 2001, IBVS, 5120

\bibitem[Kato et~al.(2004b)]{kat04v803cen}
  Kato, T., Stubbings, R., Monard, B., Butterworth, N.~D., Bolt, G., \&
  Richards, T.\ 2004b, \pasj, 56, S89

\bibitem[Kato et~al.(2004c)]{VSNET}
  Kato, T., Uemura, M., Ishioka, R., Nogami, D., Kunjaya, C., Baba, H., \&
  Yamaoka, H.\ 2004c, \pasj, 56, S1

\bibitem[Osaki et~al.(2001)]{osa01egcnc}
  Osaki, Y., Meyer, F., \& Meyer-Hofmeister, E.\ 2001, \aap, 370, 488

\bibitem[Patterson et~al.(1998)]{pat98egcnc}
  Patterson, J. {et~al.}\ 1998, \pasp, 110, 1290

\bibitem[Patterson et~al.(2000)]{pat00v803cen}
  Patterson, J., Walker, S., Kemp, J., O'Donoghue, D., Bos, M., \& Stubbings,
  R.\ 2000, \pasp, 112, 625

\bibitem[Roelofs et~al.(2004)]{roe04sdssj1240}
  Roelofs, G.~H.~A., Groot, P.~J., Steeghs, D., Nelemans, G.\ 2004,
  Rev. Mexicana Astron. Astrof., 20, 254

\bibitem[Smak(1983)]{sma83HeCV}
  Smak, J.\ 1983, Acta Astronomica, 33, 333

\bibitem[Solheim(1995)]{sol95amcvnreview}
  Solheim, J.-E.\ 1995, Baltic Astronomy, 4, 363

\bibitem[Stellingwerf(1978)]{PDM}
  Stellingwerf, R.~F.\ 1978, \apj, 224, 953

\bibitem[Tsugawa, Osaki(1997)]{tsu97amcvn}
  Tsugawa, M., \& Osaki, Y.\ 1997, \pasj, 49, 75

\bibitem[Warner(1995a)]{war95amcvn}
  Warner, B.\ 1995a, \apss, 225, 249

\bibitem[Warner(1995b)]{war95book}
  Warner, B.\ 1995b, Cataclysmic Variable Stars (Cambridge: Cambridge
  University Press)

\bibitem[Warner(1995c)]{war95amcvnproc}
  Warner, B.\ 1995c, in Cataclysmic Variables, ed. A. Bianchini, M. della Valle
  \& M. Orio (Dordrecht: Kluwer Academic Publishers), ~325

\bibitem[Wood-Vasey et~al.(2003)]{woo03sn2003avaw}
  Wood-Vasey, W.~M., Aldering, G., Nugent, P., \& Li, K.\ 2003, \iaucirc, 8077

\bibitem[Woudt, Warner(2003)]{wou03sn2003aw}
  Woudt, P.~A., \& Warner, B.\ 2003, \mnras, 345, 1266

\bibitem[Zacharias et~al.(2004)]{zac04UCAC2}
  Zacharias, N., Urban, S.~E., Zacharias, M.~I., Wycoff, G.~L., Hall, D.~M.,
  Monet, D.~G., \& Rafferty, T.~J.\ 2004, \aj, 127, 3043

\end{thebibliography}
\end{document}